\input epsf     
\documentstyle[mu94,12pt]{article}
\def\Ls{\cal L \rm}

\def\As{\cal A \rm}

\def\={\ =\ }
\title{MONTE CARLO SIMULATIONS OF MUON PRODUCTION}
\author {
Robert B. Palmer, Juan C. Gallardo, Richard C. Fernow, Ya\u{g}mur Torun  \\
Brookhaven National Laboratory\\
P. O. Box 5000, Upton, New York  11973-5000                 \vspace{.5pc} \\
David Neuffer                                               \\
CEBAF    \\
Newport News, VA, 23606                                      \vspace{.5pc}\\
David Winn                                                  \\
Fairfield University,  \\
Fairfield, CT, 06430-5195}

\begin{document}

\maketitle
\vspace{4pc}
\medskip
\aabstract
{Muon production requirements for a muon collider are presented. Production of
muons from pion decay is studied. Lithium lenses and solenoids are considered
for focussing pions from a target, and for matching the pions into a decay
channel. Pion decay channels of alternating quadrupoles and long solenoids are
compared. Monte Carlo simulations are presented for production of $\pi
\rightarrow \mu $ by protons over a wide energy range, and criteria  for
choosing the best proton energy are discussed.     }
\section{INTRODUCTION}

   The luminosity $\Ls$ of a muon collider is given by
\begin{equation}
\Ls\ =\ {{N^2 \gamma n_t f}\over{4 \pi \beta^* \epsilon_n}}
\end{equation}
where $N$ is the number of muons per bunch, $\gamma$ is the energy per beam
divided by the muon mass, $n_t$ is the effective number of turns made by the
muons before they decay, f is the repetition frequency, $\beta^*$ is the
Courant-Snyder parameter at the focus, and $\epsilon_n$ is the r.m.s.
normalized emittance of the beams (assumed symmetric in x and y). In order to
obtain a reasonable event rate, the luminosity must be proportional to the
center of mass energy squared and may be taken to be of the order of
$\Ls\geq 10^{33}\ E_{cm}^2\,cm^{-2}s^{-1}$ where $E_{cm}$ is the energy in TeV.
Possible parameters that would yield a luminosity close to this requirement,
at 4 TeV in the center of mass, are given in Table 1.

\smallskip
\smallskip
\begin{center}
\begin{tabular}{|ll|c|}
\hline
Beam energy                & TeV      &     2    \\
Beam $\gamma$              &          &   19,000 \\
Repetition rate $f$            & Hz       &    30   \\
Muons per bunch  N          & $10^{12}$  &   2     \\
Bunches of each sign       &          &   1     \\
Normalized rms emittance $\epsilon_n$  &mm mrad  &  50      \\
Average ring mag. field     & Tesla   & 6      \\
Effective turns before decay $n_t$ &       & 900     \\
$\beta^*$ at intersection   & mm     &   3      \\
\smallskip
Luminosity $\Ls$              &$cm^{-2}s^{-1}$&  $10^{35}$   \\
\hline
\end{tabular}
\end{center}
\vspace{.1in}

\begin{center}
Table 1: Parameters of a 4 TeV center of mass $\mu^{+}\mu^{-}$ Collider
\end{center}
\vspace{.1in}

   The value of the emittance $\epsilon_n$ used is limited by the technology
used to cool the beams, and is chosen here to be consistent with that
believed (1) obtainable with ionization cooling. The intersection point
(IP) $\beta^*$ will be limited by the design of the chromatic correction
system at the IP, and by the achievable bunch length. The value given here is
believed to be technically possible. The number of turns
$n_t$ is set by the average bending field in the collider ring. The value
taken here corresponds to a mean field of 6 Tesla, which is probably as high
as is technically feasible. The repetition rate $f$ is taken to be 30 Hz,
and is constrained by proton source, power consumption, and radiation
considerations.

   With the above parameters, the required luminosity is achieved, but the
number of muons per bunch $N$ is large ($2\ 10^{12}$). Stacking many smaller
muon bunches to achieve such a population would be hard because of the life
time limitations of the muons. Thus if the proton bunch population is to be
kept reasonable, the production must be efficient, i.e. we require a high
value of the number of captured muons per initial proton $\eta_\mu=n_\mu/n_p$.

   Since pion multiplicity rises with proton energy, the value of $\eta_\mu$
tends also to rise with the proton energy used. But if the energy is
allowed to rise, then the cost and energy consumption of the proton source
will also rise. Thus the requirement is for the highest number of muons per
proton $\eta_\mu$, at the lowest possible proton energy $E_p$.

\section{THEORY}

\subsection{Introduction}

We consider muons made by the decay of pions generated by the interaction of a
proton beam with a metal target. With 30 GeV protons, the pion multiplicity is
of the order of one.  Almost every pion made, if it does not interact in the
target or otherwise get lost, will decay into a muon. The potential value of
$\eta_\mu$ is thus of the order of one. At proton energies lower than 30 GeV,
the multiplicity falls, but remains substantial until below the $N^*$ resonance
(at $E_p=0.73\,GeV$). The challenge is to target efficiently and capture as
large a fraction of the pions as possible; then to capture as large a fraction
as possible of the muons from their decay.

Previous estimates (2) of the efficiency of such capture were low ($\approx
10^{-3}$), and only moderate luminosity was possible. Such estimates assumed
conventional focussing technology and decay channels with restricted momentum
acceptances ($\pm 5\%$). In this case, the values of $\eta_\mu$ are
proportional to the square of the momentum acceptances; one factor from the
fraction of pions accepted from the target, and another from the fraction of
muons accepted from the decay of those pions. Since the luminosity per pulse is
proportional to the square of the number of muons, the luminosity goes as the
fourth power of the acceptances. If these acceptances can be raised, then a
major improvement in $\eta_\mu$ may be expected.

In this paper we discuss two methods of capturing pions from the production
target: a) lithium lenses followed by a decay channel consisting of alternating
quadrupoles, and b) a high field solenoid matched adiabatically to a lower
field solenoid decay channel.

\subsection{Lithium Lenses and Quadrupole Channel}

\subsubsection{Capture from target}

    The Courant-Snyder parameter $\beta$ in a long lithium lens, assuming
uniform current density, is
 \begin{equation}
\beta \ =\ {{p\ \theta_{max}}\over{B_{max}\ c}}
 \end{equation}
where $p$ is the particle momentum in eV/c, $c$ is the velocity of
light. If $\theta_{max}$ is the maximum angular amplitude accepted, and
$B_{max}$ is the field on the surface, its radius $a$ is
 \begin{equation}
a\ =\ \beta\ \theta_{max}\ =\ {{p\ \theta_{max}^2}\over{B_{max}\ c}}
 \end{equation}
Inserting the target inside the lithium lens maximizes the yield of captured
pions. The required length $\ell_1$ of the lens is half the betatron
wavelength,  $\lambda/2$; if we express the maximum angle $\theta_{max}$ as
$\hat{p}_t/p$ then,
 \begin{equation}
\ell_1\  =\ {{\pi}\over{2}}\ {{\hat{p}_t}\over{B_{max}\ c}},
 \end{equation}
which, for a given transverse momentum acceptance, is independent of the
momentum $p$. The radius $a_1$ of the lens is
 \begin{equation}
a_1\ =\  {{\hat{p}_t^2}\over{B_{max}\ c}}\ {{1}\over{p}}
 \end{equation}
which rises as the momentum falls. For $B_{max}=10\ T$, and a nearly ideal
$\hat{p}_t=0.6\ GeV/c$, then $\ell_1=0.31\ m$. The radius will be 12 cm at 1
GeV/c, falling to 3 cm at 4 GeV/c.

   The required current $I$ is
 \begin{equation}
I\ =\ {{2\pi\ \hat{p}_t^2}\over{\mu_o\ c\ p}}
 \end{equation}
which falls as the momentum rises, independent of the surface
field. For $\hat{p}_t=0.6\ GeV/c$, the current falls from a value of 6 MA at 1
GeV/c to 1.5 MA at 4 GeV/c.

These currents and radii are far larger than those in currently used lenses (eg
FNAL: a=1 cm, I=0.5 MA), but they may still be possible. The temperature rise
and central pressures are proportional only to the surface field, and this we
kept constant. However, it is clear that this type of focussing is better
suited to the higher momenta.

  At a fixed momentum, the normalized total emittance of the resulting beam
$\epsilon_{tn}(tgt)$ is set by the target length $\ell_{tgt}$
 \begin{equation}
\epsilon_{tn}(tgt)\ = {{\ell_{tgt}}\over{2}}\ {{\hat{p}_t^2}\over{m_{\pi}\ p}}
 \end{equation}
where $m_\pi$ is the pion mass in units of $eV/c^2$ and the momenta $p$
and $\hat{p}_t$ are in units of eV/c. At 1 GeV/c,
a target length of 0.18 m (for Cu), and a maximum $\hat{p}_t$ of 0.6 GeV/c, the
emittance has a
value of 0.23 m, but has fallen to a 0.06 m at 4 GeV/c.

For a point source, with momentum spread ${dp\over p}=\delta$, the total
normalized emittance $\epsilon_{tn}(mom)$ is
 \begin{equation}
\epsilon_{tn}(mom) \= {{\pi}\over{2}}\ {{\hat{p}_t^3\ \delta}\over{B_{max}\ c\
m_{\pi}\ p}}
 \end{equation}
which also falls with momentum. This equals the emittance for target length
when
 \begin{equation}
\delta \= {{\ell_{tgt}\ B_{max}\ c}\over{\pi\ \hat{p}_t }}
 \end{equation}
which for the above parameters is $\pm 0.28.$

\subsubsection{Quadrupole channel}

   The focussing strength $k$ (defined to be the inverse of the focal length)
of a thin quadrupole lens is
 \begin{equation}
k = {{\ell_q\ B_{pole}\ c}\over a p}
 \end{equation}
where $B_{pole}$ is the pole field at the pole radius $a$, and the momentum
$p$ is in $eV/c$. If the phase advance
per cell is $\psi$, and a half cell length is $\ell_h$, then
 \begin{equation}
s \= \sin{{\psi \over 2}} \= {{\ell_h\ k}\over{2}}
 \end{equation}
The average $\beta$ is $2 \ell_h/\psi$   and the maximum $\beta$ is
 \begin{equation}
\beta_{max} \= {{2}\over{k}}\ \sqrt{{{1+s}\over{1-s}}}
 \end{equation}
Thus, the unnormalized acceptance of a quadrupole channel varies inversely
with the momentum
 \begin{equation}
{\cal A} = {  a^2 \over  \beta_{max}  } =
{a \ell_q B_{pole} c\over 2 p} \sqrt{{{1-s}\over {1+s}}}
 \end{equation}
$\As$ is zero until the momentum has risen to the value $p_{min}$
corresponding to the onset of the stop band at a phase advance per cell of
$\pi$. This minimum accepted momentum is
 \begin{equation}
p_{min} = {\ell_q^2 \over F} {B_{pole} c\over 2 a},
 \end{equation}
where $F=\ell_q/\ell_h$ is the fraction of length full of quads.
For nominal capture of 1 GeV/c pions we require a $p_{min}$ of the order of
$p/4=0.25\ GeV/c$. Then for $B_{pole}=6\ T$, $a=0.15\ m$ and $F=1/2$,
then we obtain $\ell_q=0.14\ m$. For higher momenta $\ell_q$ should be
increased as the square root of the momentum.

\begin{figure}[tbh]
\centering
\epsfxsize=7.5cm \epsfxsize=7.5cm \epsfbox{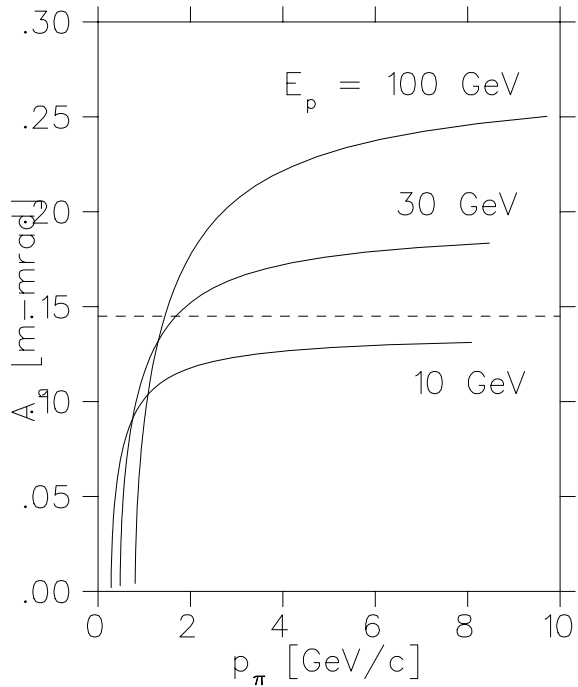}
 \caption{ Decay Channel Acceptances. Continuous lines are for quadrupole
 channels optimized for proton energies $E_p=$ 10, 30, and 100 GeV by
adjusting the quadrupole lengths; dashed line is
 acceptance for a solenoid channel with the same field.}
 \label{channel}
 \end{figure}

The unnormalized acceptance rises to a maximum at a momentum of approximately
4 times this minimum, and then falls slowly. The normalized acceptance
${\cal A}_n=\gamma\beta{\cal A}$ rises continuously, approaching an
asymptotic value of
 \begin{equation}
{\cal A}_n(quad) = {\ell_q a B_{pole} c\over 2 m_{\pi}}
 \end{equation}
Fig.~\ref{channel} shows the normalized acceptances for lattices with
quadrupole lengths $\ell_q$ of 14, 20 and 30 cm, corresponding to optimized
designs for proton energies of 10, 30, and 100 GeV. A fixed value of $F=1/2$
is assumed. It is seen that for longer quadrupoles, the maximum acceptance is
increased, but the minimum momentum rises.

\subsubsection{Matching}

   Matching from the first lens into the channel is challenging.
 The beam size after the lens is usually significantly smaller than the
aperture of the channel. An adiabatically tapered lithium lens transition
would be ideal,
providing a match at all momenta, but would involve too much beam loss from
interactions in the lithium. Instead we consider a two lens match that is
similar in function to the multiple horns used in neutrino beams (3).

\begin{figure}[tbh]
\centering
\epsfxsize=7.5cm \epsfxsize=7.5cm \epsfbox{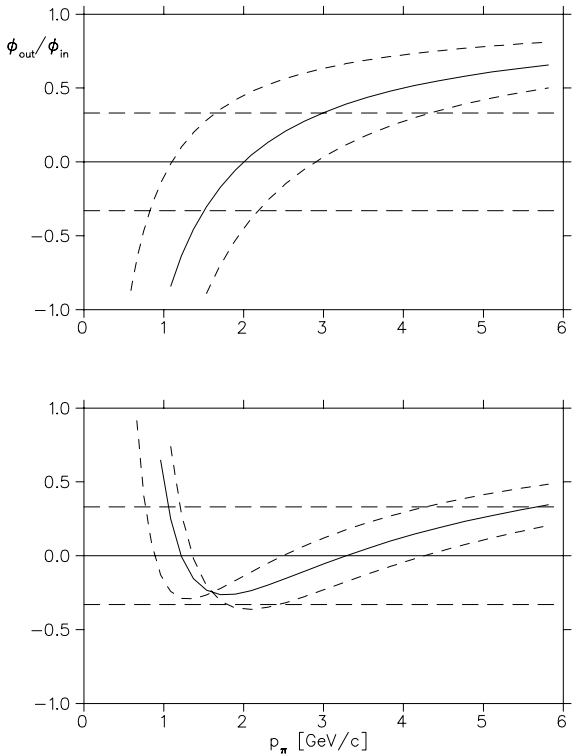}
\caption{$\theta_{out} /\theta_{in}$ vs momentum in thin lens focus systems,
solid lines are for target center, dashed lines are from target ends;
a) for single lens focussing; b) for two lens focussing.}
 \label{outin}
 \end{figure}

   We consider first a single lens focus. The solid line in
Fig.~\ref{outin}a shows the ratio of outgoing angles over incoming angles,
for a single thin lens of focal length 20 cm, set to focus 2 GeV/c momentum
particles.
$\theta_{out}/\theta_{in}$ is less than 1/3 over the momentum range
of about 1.5 GeV/c. The dotted lines show the same ratios for particles
starting at 9 cm in front or behind the nominal source, i.e. they represent
particles coming from the ends of an 18 cm target.

   Fig.~\ref{outin}b shows the same thing for a two thin lens system. The
second lens has a focal length 3 times that of the first, and both are placed
at distances from the source equal to their focal lengths. It is seen that
there are now two momenta for which the outgoing angles are zero, and the
range of momenta for which $\theta_{out}/\theta_{in}$ is less than 1/3 is now
5 GeV/c, compared with only 1.5 GeV for the single lens case. The effect of
displacing the source is to smear the distributions, but leave a net reduction
of angles over the same wide momentum range.

   Using this approach to match from the initial lithium lens of length
$\ell_1$, we place the second lens at a distance $2\ \ell_1.$ The radius
$a_2$ is set equal to that of the decay channel, and the length is
 \begin{equation}
\ell_2  \approx\ {{a_2\ p}\over{3\ \ell_1\ B_2\ c}}
 \end{equation}
which for a maximum field of $B_2=3.3\, T$ gives $\ell_2 =15\, cm$ at 1 GeV/c,
rising linearly with momentum.
\subsubsection{Parameters of Systems}
Using the above criteria we have designed lithium-lens and quadrupole channel
systems for the mean captured momenta, corresponding to initial
proton momenta of 10, 30 and 100 GeV/c.

\vspace{.1in}
\begin{center}
\begin{tabular}{|ll|ccc|}
\hline
Proton mom.  &  GeV/c   &10      &30     &100     \\
Nom. $\pi$ mom. &GeV/c  &1.3     &2.4    &5        \\
\hline
First Li Lens   &       &       &       &       \\
\hline
Surface field   &Tesla  &10     &10     &10     \\
Total length    &cm     &40     &40     &40     \\
Nominal length  &cm     &31     &31     &31     \\
Radius          &cm     &10     &5      &2.5      \\
Space           &cm     &60     &60     &60     \\
\hline
Second Li Lens  &       &       &       &       \\
\hline
Surface field   &Tesla  &3.3    &3.3    &5.7   \\
Total length          &cm     &20     &36     &40    \\
Radius          &cm     &15     &15     &15     \\
\hline
Decay channel   &       &       &       &       \\
\hline
Pole tip fields &Tesla  &6      &6      &6      \\
Quad. length    &cm     &15     &20     &30     \\
Gap length      &cm     &20     &20     &20     \\
Decay length    &m      &400    &800   &1600   \\
\hline
\end{tabular}
\end{center}
\vspace{.1in}

\begin{center}
Table 2: Lithium and Quadrupole Channel Parameters
\end{center}
\vspace{.1in}

\subsection{Solenoid Focussing}

\subsubsection{Capture from Target}

   If the capture is done inside a solenoid, then the calculations are very
simple. The required maximum radius $a_{sol}$ is
 \begin{equation}
a_{sol}\ =\ 2\ {{\hat{p}_t}\over{B_{sol}\ c}}
 \end{equation}
where $\hat{p}_t$ is the is the maximum transverse momentum in units of eV/c.
This radius is seen to be independent of momentum p. For the near ideal
$\hat{p}_t=0.6\ GeV/c$ and a conventional superconducting field
$B_{sol}=10\,$T, the radius would be 40 cm which is very large and would
correspond to a very large emittance. But hybrid solenoids can be made with
fields as high as 45 Tesla (4).  If we used a field of 33 T, then the radius
with the same assumptions, would be only 7.5 cm. At lower momenta, a somewhat
lower maximum $\hat{p}_t$ may be acceptable, for $a_{sol}=7.5\ cm$, and
$B_{sol}=28\,$T the maximum transverse momentum captured is 0.31 GeV/c.

   The solenoid
length $\ell_{sol}$ to focus pions, at a fixed momentum $p$ is
 \begin{equation}
\ell_{sol}\ ={{\pi}\over{2}}\ \beta \= \pi\ {{p}\over{B_{sol}\ c}}
 \end{equation}
which rises with momentum. For a 28 T field, the length is 37 cm for $1
\, GeV,$ rising to 150 cm for 4 GeV. But one notes that for a long solenoid,
the
beam will be captured in the solenoid at all momenta, and for all source
positions, an almost ideal situation.
In this case, the normalized acceptance of the resulting beam is
 \begin{equation}
\As_{n}(sol) = {a_{sol} \hat{p}_t\over m_\pi } = {2 p_t^2
\over B_{sol} c m_{\pi} } =
{a_{sol}^2 B_{sol} c\over 2 m_{\pi}}
 \end{equation}

In practice we cannot maintain such a solenoid over any significant length and
must match it into a lower field decay channel.

\subsubsection{Decay Channel}

   For a solenoid decay channel we chose a field of 7 Tesla, which is easily
attainable with superconducting magnets. The value is a little higher
than that taken for the quadrupole pole tip fields, since the fields seen by
the conductors in a quadrupole will be somewhat higher. We take the radius of
the channel  to be the same as that in the quadrupole case, i.e. 15 cm.

   The acceptance of this channel will be seen to be the same as that for the
7.5 cm, 28 Tesla capture solenoid. If a suitable wide band matching can be
provided, then all pions captured by the target solenoid will be transferred to
the channel, independent of their momentum.

\subsubsection{Match}

\begin{figure}[tbh]
\centering
\epsfxsize=7.5cm \epsfxsize=7.5cm \epsfbox{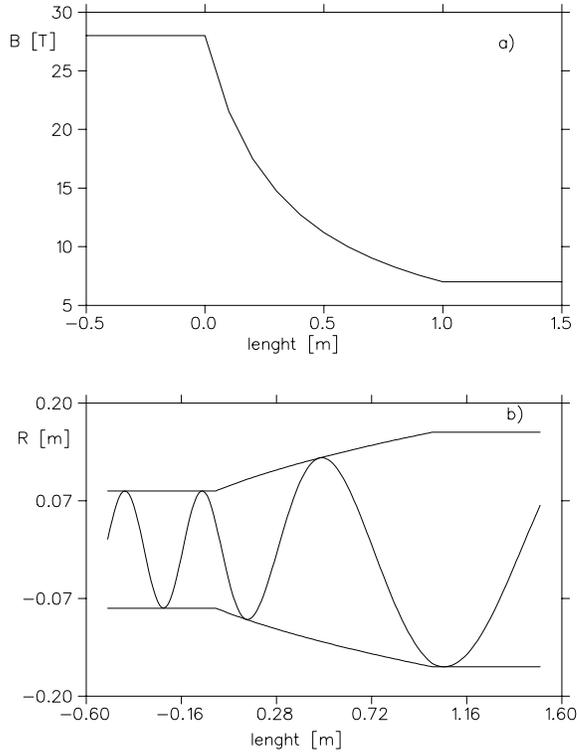}
\caption{Solenoid Matching: a) Magnetic field vs. length; b) Radius vs. length,
with a typical trajectory of a particle.}
 \label{match}
 \end{figure}

   The match between the target capture solenoid and the decay channel solenoid
can be made without significant
loss if the field and radius are varied so as to maintain the same acceptance,
and if the rate of change of $\beta$ with $z$ is
small compared with one. Defining $d\beta/dz=\epsilon$, we obtain (5)
 \begin{equation}
B \= {{B_o}\over{1\ +\ \alpha\ z}}
 \end{equation} \begin{equation}
a\= a_o\ \sqrt{1\ +\ \alpha\ z}
 \end{equation}
where
 \begin{equation}
\alpha = {cB_o  \epsilon \over 2 p}
 \end{equation}
   It is found that for values of $\epsilon$ less than $0.5$ there is
negligible loss of particles.
Fig.~\ref{match} shows the field and
dimension profiles of such a match.

\subsubsection{Parameters of Solenoid Systems}

  In table 3, parameters are given of solenoid capture and channel systems
optimized for pion momenta of 0.5, 0.9, 1.3, and 2.8 GeV/c; corresponding
approximately to the peak pion production from proton momenta of 3, 10, 30 and
100 GeV/c. These pion momenta are lower than those given for the lithium
lens-quadrupole case, see Table 2; reflecting the loss of low momenta
particles from the channel cutoff in the lithium lens-quadrupole case.

\vspace{.1in}
\begin{center}
\begin{tabular}{|ll|cccc|}
\hline
Proton mom.          &  GeV/c   &3     &10    &30     &100     \\
Nom. $\pi$ mom.      &GeV/c    &0.5     &0.9    &1.3   &2.8        \\
\hline
Target length        &cm        &24     &24     &24     &24     \\
Capture field        &Tesla     &28     &28     &28     &28     \\
Solenoid radius      &cm        &7.5    &7.5    &7.5    &7.5    \\
Transition length    & m       &0.5     &0.9     &1.3   &2.8     \\
Channel field        &Tesla    &7       &7       &7       &7       \\
Channel radius       &cm        &15     &15     &15     &15     \\
Decay Length         &m        &100     &200    &340    &500  \\
\hline
\end{tabular}
\end{center}
\vspace{.1in}

\begin{center}
Table 3: Solenoid Focus and Channel Parameters
\end{center}

\section{MONTE CARLO RESULTS}

\subsection{Monte Carlo Program}
   A Monte Carlo program has been written to give a first approximation
to the performance of the capture systems described above. The program, at
this time, contains many approximations and the results obtained from it must
be taken with some caution.
      \begin{itemize}
      \item
Pion production spectra use the Wang (6) formulation.
These distributions were derived by fitting proton proton interaction cross
sections. The $\pi$ multiplicities are slightly lower than those given by H
Boggild and T. Ferbel (7) and
significantly lower than those given by a nuclear calculation for Cu (8).
Our estimates should thus be conservative.
       \item
The pion momentum distributions given by the Wang
formula are peaked somewhat lower than those given by the nuclear calculation
for hydrogen, but higher than that given by the calculation for Cu.
Clearly, the use of the Wang formula is unsatisfactory, but the qualitative
results obtained are probably correct.
      \item
Particles were followed using the paraxial approximation. This is a reasonable
approximation for the production at 10 GeV and above, but is a poor
approximation for pions produced by protons below 3 GeV.
      \item
The program assumed that all particles are relativistic. This again is a poor
approximation for proton production below 3 GeV.
       \item
   The initial proton beam was assumed to have an rms transverse radius of 1
mm, and a divergence of 1 mrad. The target was taken to be Cu with
an interaction cross section of 0.782 barns. Pions passing through the target
were reabsorbed with a cross section 2/3 of the above, and no tertiary pion
production was included. Coulomb scattering, energy loss and straggling were
calculated from Gaussian formulae. The pion decay lifetime was taken to be
$2.603\times 10^{-8}\,s,$ the branching ratio into muons was assumed to be
100\%. The kinetic energy distribution of decay muons was taken to be flat.
Pions or muons which exceeded the aperture of focus components were assumed
lost.
      \end{itemize}

\subsection{Results of Simulations}

Table 4a gives the muon production for different initial proton energies, for
the two capture systems described above: a) lithium lenses and quadrupole
channel; b) solenoid capture and channel. The ``capture $\%$ '' given is the
fraction of pions that decay into muons that remain always within the focus
channel. These fractions, particularly in the case of solenoid focussing, are
relatively insensitive to the details of the pion production as given by the
Wang formula.  The ``$\mu/p$'' ratio is the product of the  ``capture $\%$ ''
and the average charged pion multiplicity (``$\pi$ mult.''), and gives the
final number of muons produced per initial proton. The average $\mu$ momentum
is that found in the decay channel at the end, and the ``rms mom. $\%$'' is the
rms muon momentum spread divided by the average.
\begin{figure}[tbh]
\centering
\epsfxsize=7.5cm \epsfxsize=7.5cm \epsfbox{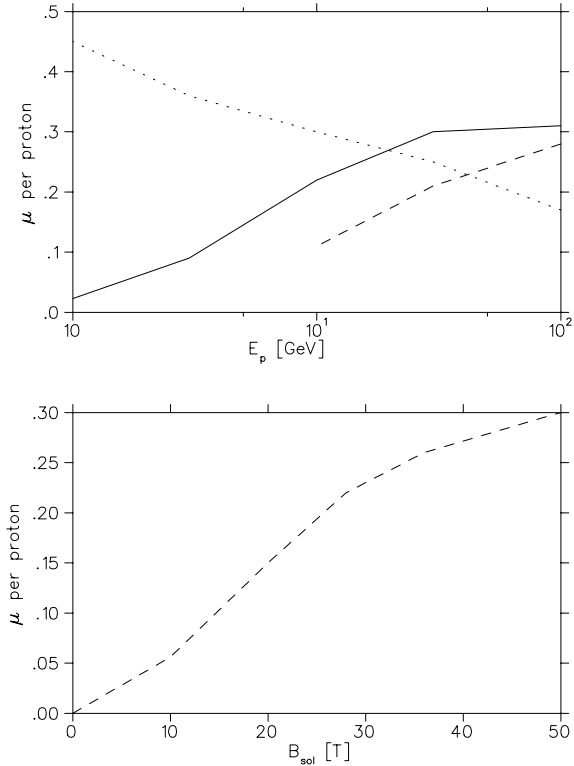}
\caption{Monte Carlo Results:   a)Muons per proton vs. proton energy; dashed
line is for lithium lens system; solid line is for all solenoid systems; dotted
line is capture efficiency for solenoid systems;    b)Muons per proton vs.
solenoid field.}
 \label{results}
 \end{figure}

Fig.~\ref{results}a shows the calculated number of muons per proton for the
different cases. It is seen that the lithium lens + quadrupole channel systems
are universally worse than the solenoid systems; although the difference is
shrinking as higher proton energies are used. In the all-solenoid cases the
production increases with proton energy, but the increase levels off above
30 GeV. The dotted line in this figure shows the capture efficiency for the
solenoid case. In reality this efficiency would be less at the lowest energies
because of production in the backward direction that is not represented by the
Wang formulation.

Table 4b and Fig.~\ref{results}b show the dependence of muon production on
the capture magnetic fields. All cases are for solenoid focussing from 10 GeV
protons. As expected, the production rises monotonically with field, but the
gain begins to saturate at fields above about 30 Tesla.

Table 4c compares capture efficiency (at 10 GeV with solenoid focussing) for a
60 cm long, 1 cm radius beryllium target with that from the shorter copper
targets assumed above. It is seen that there is little difference between them.
It must be noted however that this comparison does not include any differences
in the production multiplicities or distributions.

\vspace{.1in}
\begin{center}
\begin{tabular}{|l|cccccc|}
\hline
Method&p Energy&capture &$\pi$ mult.&$\mu/p$&ave $\mu$ mom.&rms mom. \\
      &GeV          & \%              &       &       &GeV/c         &\%   \\
\hline
Li lens +& 100       &15           &1.8        &0.28   &8             &120  \\
quads   & 30        &17           &1.2        &0.21   &2.3           &70   \\
        & 10        &16            &0.7       &0.11   &1.1           &60   \\
\hline
Solenoid&100       &17            &1.8     &0.31   &1.5           &110  \\
         &30        &25            &1.2     &0.30    &1.0           &140  \\
         &10        &30            &0.7     &0.22   &0.6           &80   \\
         &3         &35            &0.3     &0.09   &0.34          &60   \\
         &1         &46            &0.05    &0.023  &0.14          &66   \\
\hline
\end{tabular}
\end{center}

  \begin{center}
a) Dependence on proton Energy
   \end{center}

\smallskip
\smallskip
\begin{center}
\begin{tabular}{|c|c|ccc|}
\hline
Solenoid Field&Channel Field&capture &$\mu/p$&ave $\mu$ mom.      \\
   Tesla      &Tesla              &\%       &       &GeV/c              \\
\hline
20            &5                 &22       &0.15    &0.46       \\
28             &7               &30      &0.22    &0.60       \\
36             &9               &37      &0.26    &0.61       \\
\hline
\end{tabular}
\end{center}

  \begin{center}
b)  Dependence on capture Magnetic Field
   \end{center}
\smallskip
\smallskip
\begin{center}
\begin{tabular}{|l|ccccc|}
\hline
Tgt Material&Tgt length &Tgt rad.&capture &$\mu/p$&ave $\mu$ mom.      \\
               & cm            &mm        &\%       &       &GeV/c
\\
\hline
Copper      &   24             &3          &31        &0.22      &0.6      \\
Beryllium   &   60             &10         &29        &0.20      &0.6      \\
\hline
\end{tabular}
\end{center}

  \begin{center}
c)  Dependence on target materials
   \end{center}

\begin{center}
 Table 4: Muon Production from Monte Carlo Studies
\end{center}

\subsection{Choice of Proton Energy}
Using a high field solenoid for capture, we find that the capture efficiency
rises as the energy falls, at least down to a proton energy of 3 GeV.
Similarly, the number of muons made per unit of beam energy also rises, at
least down to an energy of the order of 3 GeV. Thus from the point of view of
muon production economy and efficiency, it seems desirable to use  this
relatively low energy.

But as the proton energy falls, a larger number of protons are needed to
obtain the required number of muons if we assume a single bunch targetting.
Problems might arise from targeting larger bunches, and severe space charge
problems arise in the proton ring used to bunch them prior to extraction and
targeting. The tune shift in a ring whose mean bending field is $B_{ave}$, for
a Gaussian bunch of length $\sigma_z$, is given by
 \begin{equation}
\Delta \nu\ =\ {{N\ r_o \ m_p}\over
{\sqrt{2\pi}\ 2\ \sigma_z\ \gamma_p \ \epsilon_n \ B_{ave} \ c}}
 \end{equation}
where $N$ is the number of protons in the bunch, $r_o$ is the classical radius
of the proton, $m_p$ is the proton mass in electron volts, $\gamma_p$ is the
energy of the proton divided by its mass, $\epsilon_n$ is the rms normalized
emittance of the protons and $c$ is the velocity of light.

If longitudinal emittance of the muons is to be kept as low as possible then
the proton bunch length must be less than a value, obtained from a Monte Carlo
study of phase rotation, which is proportional to the
average pion momentum,
  \begin{equation}
\sigma_z \approx 3.0\,[m]\sqrt{{1\over \gamma_p}}
  \end{equation}

\begin{figure}[tbh]
\centering
\epsfxsize=7.5cm \epsfxsize=7.5cm \epsfbox{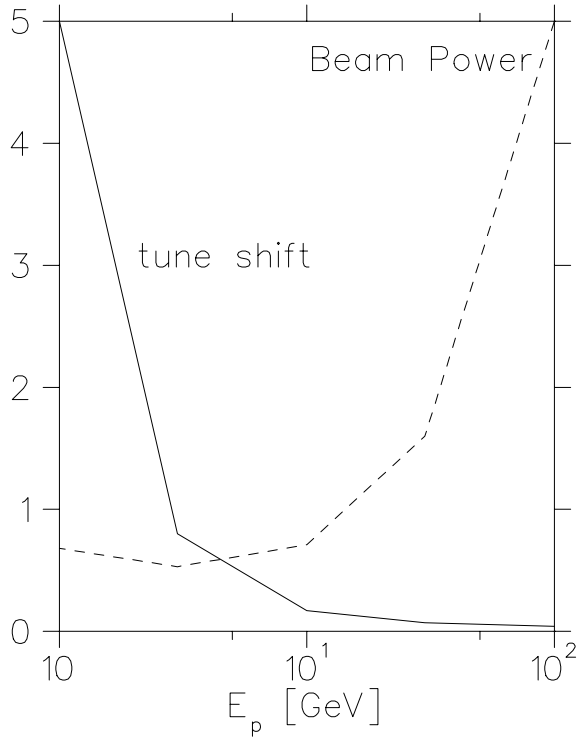}
\caption{Choice of Proton Energy: the continuous line shows the tune shift,
and the dashed line the relative proton beam power, each plotted vs. the
proton beam energy.}
 \label{choice}
 \end{figure}

For $B_{ave}$ of 4 Tesla, an rms normalized emittance of 62 mm mrad ($95\,\%$
emittance of 375 $\pi$ mm mrad), and values of N such as to give  $10^{13}$
muons using a 28 Tesla solenoid system, then we obtain the tune shifts
given in Table 5 and Fig.~\ref{choice}. In this table and figure the proton
beam power is also given.
\vspace{.1in}
\begin{center}
\begin{tabular}{|ll|ccc|}
\hline
Proton energy    &   GeV      &    3    &    10     &     30   \\
\hline
Protons required     &   $10^{13}$&2 x 11  &2 x 4.5    & 2 x 3.3    \\
Proton beam power&   MW       &   3.2   &   4.3    &    9.5    \\
Bunch len. req.  &   m       &    1.7  &    1      &     0.58 \\
Tune Shift       &            &    0.8  &    0.17   &   0.07    \\
\hline
\end{tabular}
\end{center}
\vspace{.1in}

\begin{center}
 Table 5: Proton Source parameters
\end{center}
\vspace{.1in}

Assuming a largest acceptable tune shift of about 0.2 would indicate a
preferred proton energy of 10 GeV. At this energy, the total required number of
protons (0.9 x $10^{14}$), and the required repetition rate of 30 Hz are also
close to the specification of the spallation source studied at ANL (9) and may
thus be taken as reasonable values. It is noted however, that in such rapid
cycling machines, the average bending fields are far less than the value of 4 T
assumed above. Thus, in order to bunch without excessive space charge tune
shift, we would require an additional fixed field superconducting bunching
ring. This tune-shift limit is valid for a single bunch in a circular machine.
The limit can be evaded in a linear machine or in schemes where multiple
bunches are arranged to be targetted simultaneously, for example, by using
dog-legs as ``delay-lines'', multiple rings or multiple synchronized kickers.

\section{CONCLUSION}

\subsection{Caveats}
It must be emphasized that the above calculations contain many
approximations. In particular, pion production spectra used are those given by
an approximate formula that was obtained by fitting proton-hydrogen data,
rather than proton-Cu as assumed in most cases here.
The tracking program used the paraxial approximation and assumed
that all particles were relativistic. These are poor
approximations for proton production much below 3 GeV.

\subsection{Solenoids vs. Lithium Lenses and Quadrupole Channels}
Despite the approximations used in this study, it seems clear that the use of
high field solenoids for both capture and decay channels is to be preferred
over the use of lithium lenses and quadrupole channels.
      \begin{itemize}
      \item
Given a capture solenoid approaching 30 Tesla, the absolute capture of
muons per initial proton appears higher than with any plausible lithium lens
system, at all energies. It is clearly superior for proton energies of 10 GeV
and below.
       \item
The use of a solenoid, instead of lithium lenses, allows the use of the same
target, capture and decay channel for both signs of pions. This would be a
significant saving. It may not, however, allow the use of muons of both signs
from a single proton bunch. If rf is used to bunch rotate the muons,
the polarity of this rf has to be different for the two signs.
       \item
The technology of high field solenoids is more mature than that of large
lithium lenses. Life time is less likely to be a problem, although questions
of radiation damage must be studied.
       \end{itemize}

\subsection{Choice of Proton Energy}
   The proton beam energy required per muon, falls with proton energy down to a
value around 3 GeV, but the number of protons required per bunch rises and the
tune shift problems in the proton accelerator also rise as the energy falls. A
reasonable compromise appears to be around 10 GeV.

\subsection{Proposed parameters}
On the basis of the above considerations, we propose the use of
          \begin{itemize}
          \item Solenoid capture, using a 28 Tesla 7.5 cm radius magnet.
          \item 10 GeV protons, using two bunches of $3-5\times 10^{14}$
protons.
          \item We expect at least 0.2 muons per proton, thus generating $6-10
\times 10^{12}$ muons of each sign. This appears adequate to assure final
bunches
of $2\times 10^{12}$ in the collider, yielding, with the other parameters
given in
table 1, a luminosity of $10^{35}\ cm^{-2}\ s^{-1}$.
           \end{itemize}
\section{Acknowlegement}
This research was supported by the U.S. Department of Energy under Contract No.
DE-ACO2-76-CH00016 and DE-AC03-76SF00515.

\section{REFERENCES}
\noindent 1. D. V. Neuffer, R. B. Palmer, Proc. European Particle Acc. Conf.,
London (1994)

\noindent 2. R. J. Noble, {\it  Advanced Accelerator Concepts}, AIP Conf. Proc.
{\bf 279} (1993) 949

\noindent 3. R. B. Palmer, {\it Magnetic fingers}, Proc. of the Informal
Conference on Experimental Neutrino Physics, Cern 65-32 (1965), C. Franzinetti
editor.

\noindent 4. Physics Today, Dec (1994), p21-22

\noindent 5. R. Chehab, J. Math. Phys. {\bf 5} (1978) 19.

\noindent 6. C.L. Wang, Phys. Rev. {\bf D10} (1974)  3876.

\noindent 7. H. Boggild and T. Ferbel, Annual Rev. of Nuclear Science, 24
(1990)
74

\noindent 8. D. Kahana, private communication.

\noindent 9. Yang Cho, et. al., ANL-PUB-081622, {\it Proc. Int. Collab. on
Advanced Neutron Sources}, Abbingdon, UK., May 24-28 (1993).
\end{document}